\documentclass[conference]{IEEEtran}
\usepackage{url}
\IEEEoverridecommandlockouts
\usepackage{cite}
\usepackage{amsmath,amssymb,amsfonts}
\usepackage{algorithmic}
\usepackage{graphicx}
\usepackage{textcomp}
\usepackage{xcolor}
\def\BibTeX{{\rm B\kern-.05em{\sc i\kern-.025em b}\kern-.08em
    T\kern-.1667em\lower.7ex\hbox{E}\kern-.125emX}}

\usepackage{booktabs, multirow}

\usepackage {tikz}
\usetikzlibrary{shapes,arrows,positioning,calc,matrix, positioning, intersections, arrows, arrows.meta}
\usetikzlibrary{external}

\usepackage{authblk}

\begin{document}

\title{A Knowledge-Driven Approach to Music Segmentation, Music Source Separation and Cinematic Audio Source Separation}

\author[1]{Chun-wei Ho}
\author[2]{Sabato Marco Siniscalchi}
\author[3]{Kai Li}
\author[1]{Chin-Hui Lee}
\affil[1]{Georgia Institute of Technology, USA}
\affil[2]{University of Palermo, Italy}
\affil[3]{Dolby Laboratories, China\vspace{-1.2em}}


\maketitle

\begin{abstract}


We propose a knowledge-driven, model-based approach to segmenting audio into single-category and mixed-category chunks with applications to source separation. ”Knowledge” here denotes information associated with the data, such as music scores. ”Model” here refers to tool that can be used for audio segmentation and recognition, such as hidden Markov models. In contrast to conventional learning that often relies on annotated data with given segment categories and their corresponding boundaries to guide the learning process, the proposed framework does not depend on any pre-segmented training data and learns directly from the input audio and its related knowledge sources to build all necessary models autonomously. Evaluation on simulation data shows that score-guided learning achieves very good music segmentation and separation results. Tested on movie track data for cinematic audio source separation also shows that utilizing sound category knowledge achieves better separation results than those obtained with data-driven techniques without using such information.

\end{abstract}


\section{Introduction}


Music segmentation is a task of change point detection consisting of finding temporal boundaries of meaningful events, such as instrumentation, tempo, harmony, dynamics and other musical characteristics, in continuous audio streams. This problem plays a crucial role in various downstream applications, including music source separation~\cite{Manilow2020}, music transcription~\cite{plumbley2002automatic}, remixing~\cite{pons2016remixing}, and enhancement~\cite{schaffer2022music}, etc.

Segmentation is usually accomplished by signal-based techniques. For example, an energy-based data filtering procedure was proposed in~\cite{luo2023music} using a pre-trained source separation model to generate a preliminary separation output of single- and mixed-instrument segments. Next a signal-to-noise ratio (SNR) between the mixture and the preliminary separation output is calculated. Finally, a chosen target threshold and a selected perturbation threshold are used to categorize all output segments into three classes: pseudo-target, pseudo-perturbation, and pseudo-mixture. It's worth mentioning that this method was used when training a widely used separation model, namely Band-split RNN (BSRNN)~\cite{luo2023music}, on unlabeled data. In ~\cite{Foote2000}, the authors instead proposed an approach  based on novelty detection, marking the transition between two subsequent structural parts exhibiting different properties that can be measured by self-similarity matrices (SSM)~\cite{Foote2000}.

A similar problem to music segmentation is to segment continuous speech into phonemes to build models for automatic speech recognition (ASR)~\cite{Rabiner1993}. Hidden Markov model (HMM)~\cite{Rabiner1989} is a widely-used mechanism because of its capability to simultaneously capture spectral and temporal variations in speech. Given a set of utterances and their corresponding unit transcriptions, initial models can first be established. These models can then be used to iteratively segment continuous speech into modeling units according to the given transcriptions. This is often accomplished with Viterbi decoding~\cite{Rabiner1989}, and this process is known as "forced alignment" by forcing the labels of the segment sequence to follow the given sequence of the units, e.g., words or phonemes. The given sequence of units can also be defined as a sort of ``knowledge'', which helps the segmentation process. 

In music, we do not have transcriptions as in speech, but music scores can serve as an equivalent source of "knowledge"~\cite{Moran2021, Boone2017}. Therefore, in this paper, we propose a knowledge-driven framework for music segmentation.
In traditional supervised learning, models are trained using annotated data sets, where the desired outputs, e.g., segment boundaries for meaningful events such as instrumentation and tempo in music, are explicitly provided by human experts. This annotated data serves as ground truth to guide the learning process. In contrast, the proposed knowledge-driven approach does not rely on any externally pre-segmented training data to function. This allows models to train on a more realistic scenario when segment boundaries are not available. We test our proposed knowledge-driven approach mainly on simulation data due to an availability of built-in information about segment labels and their boundaries for detailed evaluations. First we evaluate music event classification after segmentation. Next we check the effectiveness of score-guided HMM-based alignment to extract single-instrument segments from continuous audio streams.

After performing knowledge-driven segmentation, we investigate two potential applications: (i) Music source separation (MSS) using the simulated dataset described above, (ii) Cinematic source separation (CASS). Both tasks can be viewed as knowledge-driven separation methods.
Compared to the aforementioned scenario using simulated segmentation music data, cinematic audio processing (including segmentation~\cite{hung2022large} and separation~\cite{kim2023sound}) deals with more realistic and diverse audio source, including speech, music, environmental sounds, and sound effects. These data typically originate from TV shows or movie soundtracks and have important applications in the film and television industry. For example,  frame level deep-learning-based methods, such as TVSM~\cite{hung2022large}, have recently been proposed for speech and music segmentation in TV shows. Moreover, the 2023 Sound Demixing Challenge~\cite{kim2023sound} (SDX23) - also known as Divide and Remaster, DnR-v2, focused on separating speech, music, and sound effects. DnR-v3~\cite{watcharasupat2024remastering} and DnR-nonverbal~\cite{kim2023sound} later extended DnR-v2.

We compare music separation performances with widely used music separation tool packages, e.g., Demucs~\cite{defossez2019demucs} and BSRNN~\cite{luo2023music}. Finally, we evaluate CASS performances on DNR-v2 and DNR-nonverbal datasets using SepReformer~\cite{shin2024separate}. Our preliminary MSS results show the method’s effectiveness on simulated data, and the CASS results achieve state-of-the-art performance on the DNR-nonverbal dataset.


\section{Related Work}
In the past decade, music segmentation research has shifted from feature-engineered~\cite{nieto2016systematic} and probabilistic approaches~\cite{chacon2015probabilistic} toward deep, data-driven models capable of capturing both local musical cues and long-range structure. Early deep learning efforts employed convolutional~\cite{guan2018melodic} and recurrent neural networks~\cite{o2016recurrent} to learn timbral and temporal patterns directly from spectrograms, improving boundary detection over traditional self-similarity~\cite{goto2003chorus} and novelty-curve~\cite{foote2000automatic} methods. Subsequent work incorporated attention mechanisms and transformer architectures~\cite{wu2023power}, which model global musical form and repetition more effectively than RNN-based systems. Parallel advances in representation learning such as self-supervised learning~\cite{hao2025songformer, buisson2022learning} and contrastive learning~\cite{buisson2023repetition} have further strengthened segmentation performance: Representations derived from pretrained audio models robust high-level features that generalize across genres and recording conditions with minimal task-specific supervision. These systems combine the pretrained embeddings with downstream classifiers to jointly capture harmonic changes, timbral shifts, and long-range repetitions in various music recordings.

Segmentation can assist in separation. One approach is to jointly perform segmentation and separation~\cite{seetharaman2016simultaneous}, allowing models to exploit temporal structural information and improve the consistency of separated sources over time. Alternatively, segmentation results can be used for data selection: for instance, in BSRNN~\cite{luo2023music}, unlabelled audio was first segmented into single-instrument regions, which were then used for semi-supervised training, effectively increasing the quantity of usable training data while maintaining instrument-specific purity. While segmentation boundaries are often available in these tasks, weakly supervised methods~\cite{kong2020source} have also been proposed to handle cases where segment labels exist but temporal information is not provided.

A different approach involves incorporating prior knowledge, such as musical scores, instrument types, or even information about the performing artists, into building separation models. For example, score-informed methods~\cite{ewert2014score} using information from musical scores, e.g., type of instrument~\cite{8665379} and  pitch~\cite{shi1996correlations}, can help associate each separated source with its corresponding instrument. This form of knowledge-driven learning will be address later in the paper.

Cinematic Audio Source Separation (CASS) aims to decompose mixed cinematic audio into canonical stems, including dialogue (speech), music, and sound effects. While CASS differs from MSS in several aspects, certain MSS techniques can be directly applied to CASS. For instance, Solovyev et al. ~\cite{solovyev2023benchmarks} experimented Demucs4 HT, an architecture which was pre-trained on MSS and fine-tuned on CASS. Another example is BandIt~\cite{10342812}, a BSRNN-inspired architecture that achieves the state-of-the-art on the DnR-v2 dataset. Other data-driven MSS methods, such as  CrossNetunmix (XUMX)~\cite{sawata2021all} architecture, are also proven to be effective~\cite{9746005} on CASS.
One key difference between cinematic audio and music data is that the former typically lacks well-structured scores, such as MIDI files. Fortunately, in the recent Sound Demixing Challenge~\cite{kim2023sound}, the dataset includes segmentation boundaries. Recently, Hung et al.~\cite{hung2022large} proposed a speech and music segmentation model that can be used to pseudo-label cinematic audio. These segmentation boundaries can still serve as valuable knowledge to assist downstream tasks, such as source separation,  despite being less informative than music scores.

\section{Proposed Knowledge-Driven Framework}
Training a set of segmentation models typically requires a collection of single-instrument data. This can be accomplished by detecting these segments from continuous audio.
To this end, an HMM-based framework is proposed.

\subsection{Model-Based Segment Detection and Selection}
\label{subsec:HMM}

As in HMM-based recognition, not only the boundaries will vary, but also the recognized sequence will contain insertion, deletion, and substitution errors~\cite{Rabiner1989, Rabiner1993}. In contrast to ASR modeling, in which an abundance of transcribed utterances is available for building high-performance models, a major difficulty in music segmentation is a lack of single-instrument segments to train high-accuracy music classification models.
Pre-trained instrument HMMs need to be initialized and employed to identify the sequence of instrument segments. They can also be re-aligned and optimized as in the ASR modeling-building process. Most recordings come with music scores that can be processed with a similar procedure - using HMMs for forced alignment.
In order to detect and extract as many single-instrument segments as possible for model training with real-world audio, the knowledge source, such as instruments used in the music audio, are employed to improve segmentation quality with forced alignments.

Both forced alignment and recognition provide information about the active instrument segments and their corresponding boundaries, which serve as a valuable guidance for mixture-music simulation and separation model training. Other recognition models, such as DNN-HMMs~\cite{Bourlard1993, Dahl2021} and connectionist temporal classification (CTC)~\cite{graves2006connectionist} can also be adopted to produce alignments during music segmentation.

\subsection{Music Source Separation Trained with Segmented Data}
\label{sec:mss-method}
In real-world music recordings, separated tracks for individual instruments are often unavailable due to recording constraints and copyright restrictions. We address this by introducing a knowledge-driven learning framework. In this framework, all training data are constructed from identified single-instrument segments in the mixture recording. For each instrument, we first extract its corresponding single instrument segments using forced alignment. These segments are then randomly scaled, cropped, and mixed to form ``pseudo mixtures''. The separation model is trained on the pseudo mixtures and predicts the original clean segments.

We evaluate the proposed knowledge-driven separation approach in two settings: training from scratch, and fine-tuning. In former setting, the method eliminates the need for any separated tracks. In the latter, we expect the knowledge-driven strategy to provide additional performance gains for the model.

\subsection{Segment-informed CASS}
An alternative use of segmentation knowledge is to incorporate the segment information as an auxiliary input. A similar approach has been applied to singing voice separation~\cite{8665379}. We adopt the same idea for CASS to investigate whether the additional knowledge can further improve performance.

\begin{figure}
    \centering
    \includegraphics[width=\linewidth]{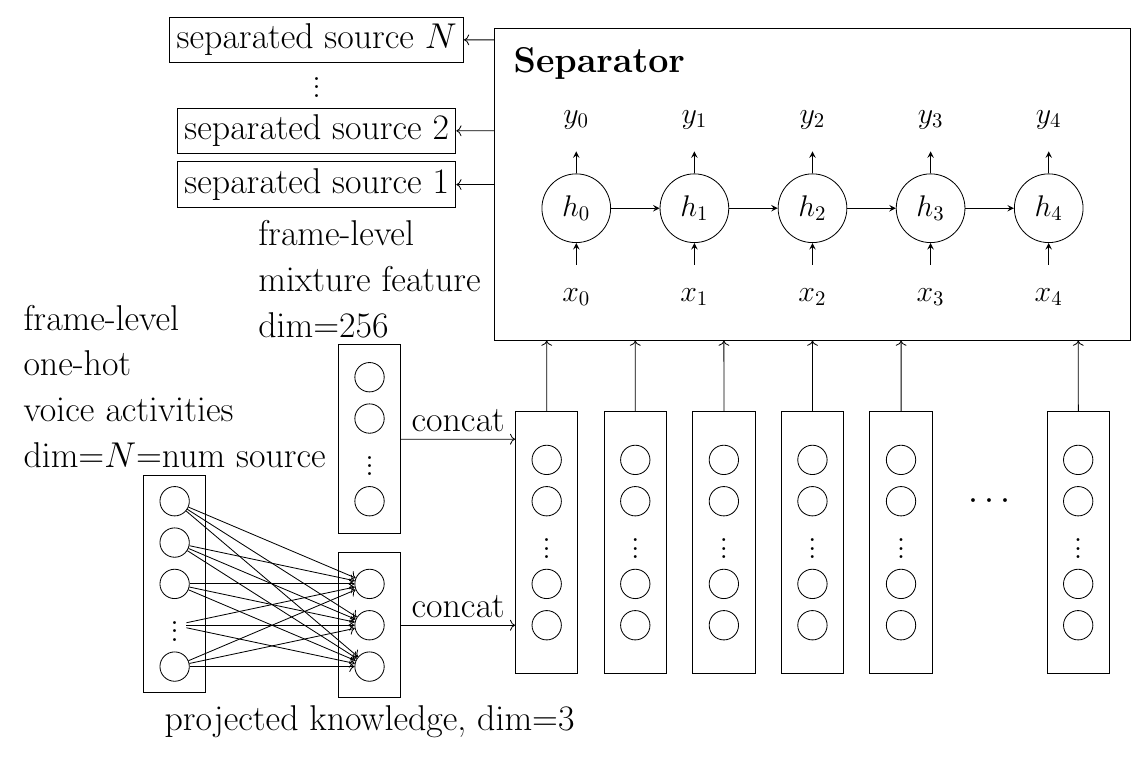}
    \caption{A knowledge-driven cinematic audio source separation system. The separator can be an RNN, a Transformer, or any separation model. It receives the mixture TF spectrogram concatenated with the projected knowledge and produces the separated spectrograms.}
    \label{fig:cass}
\end{figure}

As depicted in Figure~\ref{fig:cass}, our proposed CASS system takes as input a frame-level feature (e.g., a T-F spectrogram) and a set of frame-level voice activity indicators representing which sound sources are active. Then, the voice activity indicators are then projected into a low-dimensional embedding. Its dimension $k$ is typically the number of sound categories to be separated (dim=$k$=3 in this case). It is then concatenated with the feature vector (dim=256 here) at each frame to form the separator input. The separator can be an RNN or a Transformer. It processes the concatenated vectors and finally outputs the separated spectrograms corresponding to each target source category.

\section{Experimental Setup for Segmentaiton, Separation \& CASS}
\subsection{Data Sets and Preprocessing}
\label{sec:setup}
\subsubsection{Music Segmentation and Separation}
We conduct all music segmentation and separation experiments using Slakh2100~\cite{manilow2019cutting}, which consists of 2100 mixed tracks. To simplify the experiments, only piano and bass tracks are used. The data set is divided into three parts here: (i) Pre-train set A: a 2-instrument version of the original Slakh2100~\cite{manilow2019cutting} train set, containing 89 hours of data; (ii) Train set B (with scores): a subset (9 hours) of the 2 instrument version of the original validation set (45 hours) but annotated with music scores and segment boundaries; and (iii) Test set: a 2-instrument version of the original test set containing 11 hours of data.

Train set B is created in order to evaluate the correctness of the segmentation. In this set, the audio range from 12 to 15 seconds, consisting of solo piano, solo bass, their mixtures, and silence, with 3 to 5 seconds each. The music scores are provided as the ``knowledge''. The segmentation boundary are given as the evaluation target. This set also serves as a train set for the knowledge-driven MSS.

\subsubsection{Cinematic Audio Source Separation}
For cinematic audio source separation (CASS), we have tested our proposed framework on more realistic audio and obtained very encouraging results. In some cases, it is required to separate mixed-audio into separated sound segments, each containing a single sound source. Since category information is usually available, our proposed knowledge-driven framework, as shown in Figure~\ref{fig:cass}, can be utilized to improve performances obtained with conventional purely data-driven techniques.




For our experiments, we use DNR-v2, a dataset employed in the 2023 Cinematic Sound Demixing Challenge~\cite{kim2023sound}, for evaluation. Additionally, DNR-nonverbal~\cite{hasumi2025dnr}, an extension to DNR-v2~\cite{kim2023sound}, are used for training and evaluation.

DNR-v2 was created by randomly cropping (guided by VAD results) and scaling audio from LibriSpeech~\cite{panayotov2015librispeech}, FSD50K~\cite{fonseca2021fsd50k}, and FMA~\cite{defferrard2016fma}, and mixing them at random start times. Each clip is labeled with voice activities (i.e., start/end times for each category), speech transcription, instrument type, and sound effect (SFX). DNR-nonverbal extends DNR-v2 by incorporating nonverbal vocalizations from FSD50K (excluded in the original DNR-v2 set), including screaming, shouting, whispering, crying, sobbing and sighing, into distinct isolated categories to form sound mixtures.


\subsection{HMM-based Music Alignment Results}
For the proposed HMM-based detection and segmentation framework, we use 39-dim MFCC features~\cite{Rabiner1989}, including the first- and second-order derivatives extracted with a 25 ms window and a 10 ms shift. The music score is provided in a MIDI format. We parse each MIDI file and classify segments into one of four units: piano, bass, mixture, or silence. We then utilize Hidden Markov Model Toolkit (HTK)~\cite{young1999htk} to build and HMM for each unit. To avoid resulting in short segments, we use 300-state HMMs, where each state is characterized via a Gaussian mixture model (GMM)~\cite{Rabiner1989, sprechmann2012gaussian}.

\begin{table}[]
    \centering
    \caption{Mean absolute error (MAE) with score-informed HMM-based forced alignment. The overall MAE is 9.03 frames (with 10 ms for each frame). Detecting boundaries between mixture and either piano or bass segments is quite challenging.}
    \label{tab:forced-alignments}
    \begin{tabular}{c | c | c c c c}
        \toprule
         \multicolumn{2}{c|}{\multirow{2}{12 em}{Forced alignment segment boundary MAE (nframe)}}  & \multicolumn{4}{c}{Previous segment type} \\
         \cline{3-6}
         \multicolumn{2}{c|}{} & Silence & Piano & Mixture & Bass \\
         \midrule
         \multirow{4}{6em}{Next segment type} & Silence & - & 2.0 & 0.7 & 2.4 \\
         & Piano & 0.9 & - & 24.6 & 7.2 \\
         & Mixture & 2.4 & 29.3 & - & 14.8 \\
         & Bass & 2.5 & 6.4 & 17.1 & - \\
        \bottomrule
    \end{tabular}
\end{table}

All the training and evaluation are performed on train set~B. Our HMM-based segmentation achieves a good result, with an average MAE of 9.03 frames when compared to the reference boundaries. As summarized in Table~\ref{tab:forced-alignments}, we categorize boundaries into 12 types based on the instrument transitions. Among these types, boundaries between silence and any other segment type are the easiest for the HMMs to identify. In contrast, boundaries between mixture segments and either piano or bass segments are more challenging. Our training scheme does not require any hand-craft segmentation boundary. In fact, it only requires  music scores, which are usually available. 

In the absence of music scores, HMM-based recognition can be used to identify instrument segments, and the resulting confusion matrix is shown in Table~\ref{tab:classification}, with an overall accuracy of 94.0\%. As indicated, identifying silence and bass segments is easy. However, it tends to confuse piano with mixture segments more frequently than in other cases. With the absence of music scores, this types of segment confusions will provide false information for the next stage separation, which wouldn't happen when music scores are provided.

\begin{table}[]
    \centering
    \caption{The confusion matrix for HMM based recognition in the absence of music scores. The model tends to confuse piano and mixture segments more often than other classes.}
    \label{tab:classification}
    \begin{tabular}{c | c | c c c c}
        \toprule
         \multicolumn{2}{c|}{\multirow{2}{12 em}{Confusion Matrix of the HMM segment recognition}}  & \multicolumn{4}{c}{Reference segment type} \\
         \cline{3-6}
         \multicolumn{2}{c|}{} & Silence & Piano & Mixture & Bass \\
         \midrule
         \multirow{4}{6em}{Recognized segment type} & Silence & 2000 & 3 & 1 & 16 \\
         & Piano & 0 & 1742 & 86 & 14 \\
         & Mixture & 0 & 247 & 1834 & 23 \\
         & Bass & 0 & 1 & 79 & 1947 \\
        \bottomrule
    \end{tabular}
\end{table}

\subsection{Music Source Separation Results}


BSRNN~\cite{luo2023music} and HTDemucs~\cite{rouard2023hybrid}, are adopted as our benchmark separation models. At a sampling rate of 44,100 Hz, STFT spectrograms, computed with a Hamming window size of 4096 samples, and a hop size of 1024 samples.


To prepare the ``pseudo mixtures'' mentioned in section \ref{sec:mss-method}, we first remove all detected single-instrument segments shorter than 3 seconds. All remaining segments are next randomly cropped to a fixed duration of 3 seconds. The piano solo (target) and bass solo (perturbation) segments are then randomly mixed to create training examples. We apply data augmentation techniques following the approach described in~\cite{uhlich2017improving}, including random swapping of left and right channels for each instrument, random scaling of amplitudes within a range of $\pm 10$ dB, and a random dropping rate of 10\% of either target or perturbation segments to simulate cases where only one source is present.

We evaluate SDR~\cite{SDR2006} results, in dB, in three training scenarios, namely: (1) pre-train (PT) on train set A; (2) Knowledge driven training from scratch (KD-FS) on the train set B; (3) Knowledge-driven fine-tuning (KD-FT) from the PT model using train set B. Results are given in Table~\ref{tab:MSS}.

Table~\ref{tab:MSS} is organized into two sections, each corresponding to a widely used MSS model: BSRNN~\cite{luo2023music} and HTDemucs~\cite{rouard2023hybrid}. In both sections, KD-FS outperforms PT even with substantially less training data. This indicates that carefully selecting training data based on ``knowledge'' can enhance performance, and a large dataset is not necessary if the segmentation model (HMM in our case) is appropriate. Furthermore, in scenarios where a large amount of pretraining data is available, the KD-FT setup can be applied. Here, knowledge-based data selection can further boost the PT model’s performance, achieving better results than KD-FS due to training on a larger dataset. This trend is consistent for both BSRNN and HTDemucs, suggesting that the performance gains stem primarily from the selected data rather than the model architecture.

The last row of Table \ref{tab:CASS} presents the scenario in which the reference segmentation is used, representing the upper bound of our knowledge-driven approach under perfect boundary detection. This indicates that the method’s performance could improve further if the segmentation quality itself is enhanced


\begin{table}[]
    \centering
    \caption{The evaluation results of MSS. Knowledge-driven indicates the usage of a music score to run the forced alignment on train set B, and use the forced alignment results to select data to train/fine-tune the separation model.}
    \label{tab:MSS}
    \begin{tabular}{p{.55\columnwidth} c c}
        \toprule
         Scenario & Model & SDR (dB)\\
         \midrule
         Pre-train (PT) & BSRNN & 15.06 \\
         Knowledge-driven from-scratch (KD-FS) & BSRNN & 17.89 \\
         Knowledge-driven fine-tune (KD-FT) & BSRNN & \textbf{18.52} \\
         KD-FT with reference alignments & BSRNN & \underline{22.30} \\
         \hline
         Pre-train (PT) & HTDemucs & 14.34 \\
         Knowledge-driven from-scratch (KD-FS) & HTDemucs & 17.84 \\
         Knowledge-driven fine-tune (KD-FT) & HTDemucs & \textbf{17.90} \\
         KD-FT with reference alignments & HTDemucs & \underline{18.36} \\
         \bottomrule
    \end{tabular}
\end{table}

\subsection{Cinematic AudioSource Separation Results}
SepReformer~\cite{shin2024separate} is adopted as separator. It is a deep spectrum regressor~\cite{Du2016regression} with a separation encoder and a reconstruction decoder arranged in a temporally multi-scale Transformer U-Net. Separation is performed at the U-Net’s bottleneck. The decoder is shared across all separated sources.

Table~\ref{tab:CASS} summarizes the results on DNR-nonverbal. We compared performances with and without the inclusion of voice activity information. The model with voice activity information (4th row) significantly outperforms the version without it (3rd row) in all categories. Moreover, our model surpasses the current state-of-the-art on DNR-nonverbal (2nd row), demonstrating the strong potential of our approach. The key comparison lies between the third and fourth rows.  This suggests that, for cinematic mixtures with diverse and overlapping acoustic events, conventional separation models may have difficulties to correctly associate spectral energy to appropriate source without additional context. In contrast, when we integrate the projected voice activity representation (fourth row), the model achieves large and consistent improvements across all three categories. The gains are most pronounced for speech and sound effects, where we observe improvements of 3.35 and 4.26 dBs, respectively, over the no-activity variant. These improvements also surpass the state-of-the-art BSRNN baseline~\cite{hasumi2025dnr}, demonstrating the importance of explicitly conditioning the separator on high-level knowledge at each frame. This in turn suggests  that the proposed source-related knowledge-driven framework helps the model better allocate acoustic energy to the correct outputs, particularly in complex cinematic scenes where multiple nonverbal sound elements co-occur.



\begin{table}[]
    \centering
    \caption{SDR evaluation results on DNR-nonverbal with and without voice activity information.}
    \label{tab:CASS}
    \begin{tabular}{p{.55\columnwidth} | c c c}
        \toprule
         \multirow{2}{*}{Method} & \multicolumn{3}{c}{SDR (dB)} \\\cline{2-4}
         & Speech & Music & SFX \\
         \toprule
         BSRNN trained on DNR-v2~\cite{hasumi2025dnr} & 5.62 & 4.33 & 2.54 \\
         BSRNN trained on DNR-nonverbal~\cite{hasumi2025dnr} & 9.30 & 4.79 & 5.23 \\
         SepReformer trained w/o voice activity & 7.68 & 2.94 & 2.41 \\
         SepReformer trained w/ voice activity & \textbf{11.03} & \textbf{5.12} & \textbf{6.67} \\
         \bottomrule
    \end{tabular}
\end{table}

Next, we examine the effectiveness of utilizing category information through visualization with knowledge projection, referring to the bottom left portion of Figure \ref{fig:cass}. In the DNR-nonverbal data set, the category for each frame is labeled as speech, music, or sound effect. Speech is further divided into dialog and nonverbal and sound effects are split into foreground and background. Thus, each frame can contain dialog, nonverbal, music, foreground SFX, background SFX, or any arbitrary mixture of them, resulting in $2^5=32$ possible combinations. We iterate over all 32 possible combinations to construct 32 seven-dimensional binary voice activity vectors. Each dimension corresponds to one of the following categories: speech, dialogue, non-verbal, music, sound effect (sfx), foreground sfx, and background sfx, where 0 indicates the category is inactive and 1 indicates it is active. We then feed all 32 voice activity vectors through the knowledge projector mentioned in Figure~\ref{fig:cass}, and visualize the resulting projected 3-D representations and colored according to their category attributes.

\begin{figure}
    \centering
    \includegraphics[width=\linewidth]{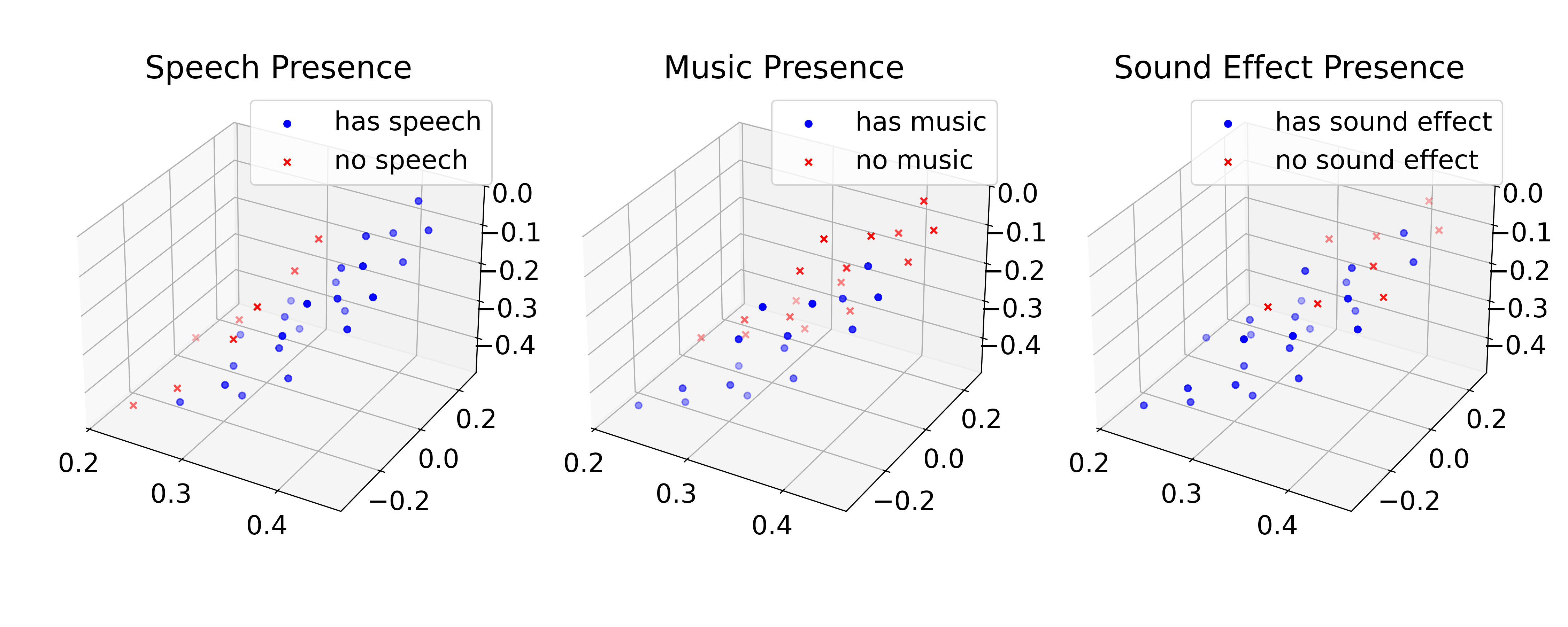}
    \caption{Grouping of the 3-D knowledge projection for all possible voice activities in the DNR-nonverbal set for (1) left: speech vs. non-speech; (2) middle: with vs. w/o music; and (3) right: with vs. w/o sound effect.}
    \label{fig:knowledge-projection}
\end{figure}

As shown in the left part of Figure \ref{fig:knowledge-projection}, the resulting 3-D feature projection successfully groups those frames containing speech from those without speech. A similar pattern is observed for music and sound effects shown in the middle and right parts of Figure \ref{fig:knowledge-projection}, respectively. Moreover, the clusters associated with speech differ noticeably from those associated with music and sound effects, whereas the clustering patterns for music and sound effects are somewhat similar. This suggests that music and sound effects are less distinguishable from each other when compared to the strong contrast between speech and the other two categories.

In addition, to illustrate the separation performance, we include a demo slide as supplemental material in the submission. In this slide, we select one example from the DNR-nonverbal set, apply our trained knowledge-driven model, and present all three separated components, speech (dialog and nonverbal), sound effects (foreground and background), and music.

\section{Summary and Future Work}
So far, we have evaluated our proposed framework on both MSS and CASS. Our preliminary results are quite encouraging for the proposed knowledge-driven learning paradigm. In fact, we have verified that the use of music scores helps detecting single-instrument segments. The detected single instrument segments can further facilitate the training process, thereby enhancing separation performance.

In addition to training data selection, we found that providing segmentation boundaries as part of the model input significantly improves performance. Moreover, the learned knowledge projection reveals clear phonetic similarities among different sound categories.

In future work, we intend to experiment on real-world continuous audio recordings. We need to improve HMM-based detection of single-instrument segments with DNN-HMMs~\cite{Bourlard1993}. When the available audio duration is very limited, e.g., in a single song of less than 5 minutes in length, pre-trained models are required. We will investigate ways to collect music materials with instrument data close to the target audio to learn appropriate pre-trained models. Once pre-trained models are used, knowledge transfer to target audio will become a key challenge. Recent adaptation techniques, applied to attention mechanisms~\cite{Subakan2021} and latent variables~\cite{Hu2025}, can also be studied.


\pagebreak
\bibliographystyle{IEEEbib}
\bibliography{icme2025references}

\end{document}